\documentclass[twocolumn,twoside]{IEEEtran}
\usepackage{mathpazo}
\usepackage{times}
\usepackage{amsmath}
\usepackage{amsfonts}
\usepackage{latexsym}
\usepackage{amssymb}
\usepackage{cite}

\usepackage{upref}
\usepackage{theorem}
\usepackage{graphicx}
\usepackage{psfrag}

\usepackage[usenames]{color}

\theoremstyle{plain}
\theorembodyfont{\normalfont\slshape}

\newtheorem{thm}{Theorem$\!$}
\newenvironment{theorem}
{\begin{thm}\hspace*{-1ex}{\bf.}}{\end{thm}}

\newtheorem{lem}[thm]{Lemma$\!$}
\newenvironment{lemma}{\begin{lem}\hspace*{-1ex}{\bf.}}{\end{lem}}

\newtheorem{prop}[thm]{Proposition$\!$}

\newtheorem{cor}[thm]{Corollary$\!$}

\newtheorem{defn}[thm]{Definition$\!$}
\newenvironment{definition}{\begin{defn}\hspace*{-1ex}{\bf.}}{\end{defn}}

\newtheorem{xmpl}[thm]{Example$\!$}

\newtheorem{cnstr}{Construction$\!$}

\newcounter{enumrom}
\renewcommand{\theenumrom}{(\roman{enumrom})}

\makeatletter
\renewcommand{\@endtheorem}{\endtrivlist}
\makeatother

\makeatletter
\renewcommand{\thefigure}{{\@arabic\c@figure}}
\renewcommand{\fnum@figure}{{\bf Figure\,\thefigure}}
\makeatother

\newcommand{\cC}{\mathcal{C}}

\newcommand{\cR}{\mathcal{R}}

\newcommand{\e}{\varepsilon}

\newcommand{\mathset}[1]{\left\{#1\right\}}
\newcommand{\abs}[1]{\left|#1\right|}

\newcommand{\sparenv}[1]{\left[ #1 \right]}

\newcommand{\be}[1]{\begin{equation}\label{#1}}
\newcommand{\ee}{\end{equation}}

\renewcommand{\leq}{\leqslant}

\renewcommand{\geq}{\geqslant}

\renewcommand{\Bbb}{\mathbb}

\newcommand{\Cref}[1]{Co\-ro\-lla\-ry\,\ref{#1}}

\renewcommand{\Bbb}{\mathbb}

\newcommand{\Fq}{{{\Bbb F}}_{\!q}}

\newcommand{\eqdef}{\triangleq}
\newcommand{\dc}{d_{\mathrm{cyc}}}

\outer\def\proclaim #1. #2\par{\medbreak
 \noindent{\bf#1.\enspace}{\sl#2\par}%
 \ifdim\lastskip<\medskipamount \removelastskip\penalty55\medskip\fi}

\begin{document}


\IEEEoverridecommandlockouts \title{\textbf{Non-linear Cyclic Codes
    that Attain the Gilbert-Varshamov Bound}}

\author{\large
Ishay Haviv,~\IEEEmembership{Member,~IEEE},
Michael~Langberg,~\IEEEmembership{Senior Member,~IEEE},
Moshe~Schwartz,~\IEEEmembership{Senior Member,~IEEE},
and Eitan~Yaakobi~\IEEEmembership{Member,~IEEE}%
\thanks{Ishay Haviv is with the School of Computer Science
at The Academic College of Tel Aviv-Yaffo, Tel Aviv 61083, Israel.}
\thanks{Michael Langberg is with the Department
  of Electrical Engineering, State University of New-York at Buffalo, Buffalo, NY 14260, USA (e-mail: mikel@buffalo.edu).}%
\thanks{Moshe Schwartz is with the Department
   of Electrical and Computer Engineering, Ben-Gurion University of the Negev,
   Beer Sheva 8410501, Israel
   (e-mail: schwartz@ee.bgu.ac.il).}%
\thanks{Eitan Yaakobi is with the Department
  of Computer Science, Technion -- Israel Institute of Technology, Haifa 32000, Israel (e-mail: yaakobi@cs.technion.ac.il).}%
\thanks{Authors appear in alphabetical order. Research supported in part by NSF grant 1526771, the Israel Science Foundation (ISF) grants No.~130/14, and No.~1624/14.}
}

\maketitle

\begin{abstract}
We prove that there exist non-linear binary cyclic
codes that attain the
Gilbert-Varshamov bound.
\end{abstract}

\begin{IEEEkeywords}
  Cyclic codes, non-linear codes, Gilbert-Varshamov bound, good family
  of codes
\end{IEEEkeywords}




\section{Introduction}
\label{sec:intro}

For a finite field $\Fq$, a \emph{cyclic code} of length $n$ is a
linear subspace $C\subseteq\Fq^n$ that is closed under cyclic
permutations, i.e., for every codeword $x \in C$, all cyclic
permutations of $x$ are also included in $C$. Cyclic codes have been
extensively studied over the last decades exhibiting a rich algebraic
structure with immense applications in storage and communication,
e.g.,
\cite{chien1964cyclic,forney1965decoding,massey1969shift,macwilliams1977theory}.

A code over alphabet $\Sigma$ is said to have \emph{(normalized) minimum distance} at least
$\delta \in [0,1]$ if any two distinct codewords in $C \subseteq \Sigma^n$ are of Hamming
distance at least $\delta n$. It is also said to have \emph{rate}
$R(C) \eqdef \frac{1}{n}\log_{|\Sigma|}\abs{C}$. Of particular interest are
\emph{good families of codes}, which are sequences of codes
$C_1,C_2,\dots$, with $C_i$ of length $n_i$, rate $R_i$, and minimum
normalized distance $\delta_i$, such that simultaneously,
\[ \lim_{i\to\infty}n_i=\infty, \qquad \lim_{i\to\infty}R_i >0,\mbox{~~~and} \qquad \lim_{i\to\infty}\delta_i > 0.\]

One of the most fundamental challenges in coding theory is to construct good families of codes.
Several such families were presented in the literature over the years. This includes, for example,
the Gilbert-Varshamov (GV) codes \cite{gilbert1952comparison,gilbert1957comparison} and the 
algebraic constructions of Justesen \cite{Justesen} and Goppa \cite{Goppa} (see also \cite{macwilliams1977theory}).
However, the existence of a good family
of \emph{cyclic} (linear) codes is a long-standing open problem, see e.g.,
\cite{martinez2006class,bazzi2006some,ding2014binary,dougherty2015open}.


In an attempt to shed some light on the problem of good cyclic codes,
some works considered close variants to cyclic codes. Quasi-cyclic and
double-circulant codes are one example, created by interleaving cyclic
codes. These families were shown to contain families of good codes
\cite[Ch.~16]{macwilliams1977theory}. Another such example is the
family of module skew codes, which are almost cyclic except for a
slight twist in the permutation. These were recently shown to contain
a good family of codes \cite{alahmadi2016long}.

In this work, we study another variant of cyclic codes --
\emph{non-linear} cyclic codes\footnote{It is common in the literature
  to define cyclic codes as linear. Thus, we shall make it a point to
  emphasize the fact that the codes we consider \emph{may be}
  non-linear by naming them \emph{non-linear cyclic codes}.}. For
simplicity of presentation we consider only the binary case. We will
show that this family of codes contains a good family which asymptotically meets
the GV bound \cite{gilbert1952comparison,gilbert1957comparison}, i.e., of
normalized distance $\delta$ and asymptotic rate approaching
$1-H(\delta)$, where $H$ stands for the binary entropy function. This
matches the best known lower bound for binary codes of minimum normalized distance $\delta$ that are not
necessarily cyclic. To the best of our knowledge, good families of non-linear
binary cyclic codes have not been previously presented in the
literature.

Our construction of good binary non-linear cyclic codes is
conceptually very simple and includes two steps.  In the first step we
construct a high-rate binary code which we call \emph{auto-cyclic}.
(All formal definitions are given below in Section~\ref{sec:pre}.)  An
auto-cyclic code is a non-linear cyclic code in which 
the set of cyclic permutations of any given codeword is of (normalized) minimum distance at least $\delta$. 
In this context, we refer to $\delta$ as the
\emph{auto-cyclic} distance.  Auto-cyclic codes are reminiscent of
\emph{orthogonal} or low \emph{auto-correlation} codes,
e.g. \cite{sarwate1979bounds,chung1989optical}.  Using a probabilistic
argument, we show the existence of a subset of $\mathset{0,1}^n$ of
asymptotic rate $1$ which is auto-cyclic with auto-cyclic distance
$\delta$ arbitrarily close to $\frac{1}{2}$.  Once a high-rate
auto-cyclic code is established, we greedily remove some of its
elements (using a slight variant of the well known greedy process that
leads to the GV bound) to obtain the desired non-linear cyclic code
$C$ of rate $1-H(\delta)$ and minimum distance $\delta$.

The remainder of this note is structured as follows.  In
Section~\ref{sec:pre} we present our formal definitions.  In
Section~\ref{sec:GV} we prove the existence of binary non-linear
cyclic codes that meet the GV bound.  Section~\ref{sec:conc} includes
concluding remarks and open questions.

\section{Preliminaries}
\label{sec:pre}

Let $[n]\eqdef\mathset{0,1,\dots,n-1}$. In the context of indices, all
addition and multiplication operations are done modulo $n$.

\begin{definition}
Let $x=x_0,\dots,x_{n-1} \in \mathset{0,1}^n$. For all $i \in [n]$,
the cyclic shift of $x$ $i$-locations to the left is defined as
\[
E^i(x) \eqdef  x_{i},\dots,x_{n-1},x_0,\dots,x_{i-1}.
\]
\end{definition}

\begin{definition}
Let $x=x_0,\dots,x_{n-1} ,y=y_0,\dots,y_{n-1} \in \mathset{0,1}^n$.
The (normalized) Hamming distance between $x$ and $y$ is defined as
\[
d(x,y) \eqdef \frac{\abs{\mathset{i \in [n] : x_i \ne y_i}}}{n}.
\]
The cyclic Hamming-distance between $x$ and $y$ is defined as
\[
\dc(x,y) \eqdef \min_{i \in [n]} d(E^i(x),y) = \min_{i \in [n]} d(x,E^i(y)).
\]
The auto-cyclic Hamming-distance between $x$ and itself is defined as
\[
\dc^*(x,x) \eqdef \min_{i: E^i(x) \ne x} d(E^i(x),x).
\]
Notice that in the definition of the auto-cyclic distance we only
consider shifts $E^i(x)$ that differ from $x$. For the all-0 vector $0^n$ and the all-1 vector $1^n$ we define the auto-cyclic distance to be $n$.  
\end{definition}

\begin{definition}
A subset $C \subseteq \mathset{0,1}^n$ is \emph{cyclic} if for every
$x \in C$ and every $i \in [n]$ it holds that $E^i(x) \in C$.
\end{definition}

\begin{definition}
We say that $C \subseteq \mathset{0,1}^n$ is an $[n,\delta]$ binary
\emph{auto-cyclic code} if $C$ is cyclic and in addition for every $x
\in C$ it holds that $\dc^*(x,x) \geq \delta$.
\end{definition}

\begin{definition}
We say that $C \subseteq \mathset{0,1}^n$ is an $[n,\delta]$ binary
non-linear \emph{cyclic code} if $C$ is cyclic and in addition, for
every $x,y \in C$, $x\neq y$, it holds that $\dc(x,y) \geq \delta$.
\end{definition}

\begin{definition}
The rate of a subset $C \subseteq \mathset{0,1}^n$ is defined by
\[
R(C)\eqdef\frac{\log_2{\abs{C}}}{n}.
\]
\end{definition}

\begin{definition}
The asymptotic rate of an infinite sequence of codes
$\cC=\mathset{C_i}_{i=1}^{\infty}$, where $C_i \subseteq
\mathset{0,1}^{n_i}$, $n_i<n_{i+1}$ for all $i$, is defined as
\[
\cR(\cC)\eqdef\limsup_{i \to \infty}R(C_i)=\limsup_{i \rightarrow \infty}\frac{\log_2{\abs{C_i}}}{n_i}.
\]
\end{definition}


\section{Non-linear Cyclic Codes}
\label{sec:GV}

We start with the following lemma that provides a probabilistic
construction of binary auto-cyclic codes of high rate.

\begin{lemma}
\label{lemma:auto}
For every $0 \leq \delta < \frac{1}{2}$ and every sufficiently large
prime $n$, there exists an $[n,\delta]$ binary auto-cyclic code of rate
at least $1-\frac{1}{n}$.
In particular, there exists a sequence
of binary auto-cyclic codes of normalized minimum distance $\delta$
and asymptotic rate $1$.
\end{lemma}

\begin{IEEEproof}
Let $n$ be a sufficiently large prime. Consider choosing a random
element $x=x_0,\dots,x_{n-1}$ in $\mathset{0,1}^n$ from the uniform
distribution.  Namely, each entry of $x$ is chosen i.i.d.~uniformly
over $\mathset{0,1}$.  We study the probability that $\dc^*(x,x) \geq
\delta$.  Let $i \in [n] \setminus \mathset{0}$.  We first analyze the
probability
\[
\Pr_x\sparenv{d(E^i(x),x)\geq \delta}.
\]
In our analysis we will use the fact that for a prime $n$, the
sequence $0,i,2i,3i,\dots, (n-1)i$ consists of elements that are all
distinct modulo $n$.  For $k \in [n]$, let $A_k$ be the indicator of
the event that the $ki$-th coordinate (modulo $n$) of $x$ differs from
that of $E^i(x)$, namely that $x_{ki}$ differs from $x_{(k+1)i}$.  As
$n$ is prime it holds that
\[
\Pr_x\sparenv{d(E^i(x),x)\geq \delta} = \Pr_x\sparenv{\sum_{k\in[n]} A_k \geq \delta n}.
\]

Let $a=a_0,\dots,a_{n-1} \in \mathset{0,1}^n$ be an arbitrary vector,
and consider the probability
\begin{multline*}
  \Pr\sparenv{\forall k: A_k=a_k}  \\
  = \prod_{k=0}^{n-1}\Pr\sparenv{A_k=a_k \mid A_0=a_0, \dots,A_{k-1}=a_{k-1}}.
\end{multline*}
Notice that each event $A_k$ depends only on $x_{ki}$ and
$x_{(k+1)i}$.  Using the fact that the sequence $0,i,2i,3i,\dots,
(n-1)i$ consists of elements that are all distinct modulo $n$, we
conclude that for $k \in \mathset{0,\dots, n-2}$:
\[
\Pr\sparenv{A_k=a_k \mid A_0=a_0, \dots,A_{k-1}=a_{k-1}} = \frac{1}{2}.
\]
Here, we used the fact that each entry $x_i$ of $x$ is uniform over
$\mathset{0,1}$ and that $x_{(k+1)i}$ is independent of the entries
$\{x_{\ell i}\}_{\ell \in [k+1]}$ that determine the events
$A_0,\dots,A_{k-1}$.  Thus,
\[
\Pr\sparenv{\forall k: A_k=a_k}  \leq \frac{1}{2^{n-1}}.
\]

We now conclude that for any $i\in[n]\setminus\mathset{0}$,
\begin{align*}
\Pr_x\sparenv{d(E^i(x),x) < \delta} & = \Pr_x\sparenv{\sum_{k\in[n]} A_k < \delta n} \\
& \leq \sum_{\substack{a \in \mathset{0,1}^n \\ d(a,0)< \delta}}\Pr_x\sparenv{\forall k: A_k=a_k} \\
&\leq \frac{2^{H(\delta) n}}{2^{n-1}}.
\end{align*}
Above, we upper bound the Hamming ball of radius $\delta n$ by
$2^{H(\delta) n}$, where $H$ stands for the binary entropy function.
By the union bound over $i \in [n] \setminus \mathset{0}$, we have
\begin{equation}
\label{eq:auto}
\Pr_x\sparenv{\dc^*(x,x)< \delta} \leq 2(n-1)\cdot2^{(H(\delta) -1)n}.
\end{equation}

Finally, define
\[C \eqdef \mathset{ x\in\mathset{0,1}^n : \dc^*(x,x)\geq \delta}.\]
By \eqref{eq:auto} we have
\[\abs{C} \geq 2^n-2(n-1)\cdot 2^{H(\delta) n} \geq
2^{n-1},\]
for $\delta<\frac{1}{2}$ and a sufficiently large $n$,
so the rate of $C$ is at least $1-1/n$. 
In addition, if $x \in C$ then any cyclic shift of $x$ is also in $C$.
\end{IEEEproof}

Equipped with Lemma~\ref{lemma:auto}, we are ready to prove our main
result stated below.

\begin{theorem}
For every $0 < \delta < \frac{1}{2}$ and $R<1-H(\delta)$ there exists a sequence of
binary non-linear cyclic codes of normalized minimum distance at least $\delta$ and 
asymptotic rate at least $R$.
\end{theorem}

\begin{IEEEproof}
Fix $\e >0$ and let $R\eqdef 1-H(\delta) -\e$.  We construct for every sufficiently large prime $n$ an $[n,\delta]$ binary non-linear cyclic code
$C$ of rate at least $R$.  Our construction has two steps.  In the first
step, using Lemma~\ref{lemma:auto}, we construct an $[n,\delta]$
auto-cyclic code $C'$ of rate at least $1-\frac{\e}{2}$.

In the second step we construct $C$ by a greedy procedure similar to
that used in the standard GV bound.  Specifically, we start with
$C=\emptyset$.  Let $x$ be any word in $C'$.  Add $x$ and all its
cyclic shifts $\mathset{E^i(x)}_{i \in [n]}$ to $C$ and remove them
from $C'$.  In addition, remove from $C'$ all words $y$ for which
$\dc(x,y) < \delta $.  In this process, since $\dc(x,y)=\min_i
d(E^i(x),y)$, we remove at most $n2^{H(\delta) n}$ words from $C'$.
Here, as before, we upper bound the Hamming ball of radius $\delta n$
by $2^{H(\delta) n}$.  Note that for any $y$ removed, we also remove
all its cyclic shifts.

We now continue in iterations, in each iteration we add an element $x
\in C'$ and all its cyclic shifts to $C$ and remove all $y$ for which
$\dc(x,y) < \delta $ from $C'$ (including $x$ and its cyclic shifts).
We continue in this fashion until $C'$ is empty.

It follows that the code $C$ constructed above
has size at least
\[\abs{C}\geq \frac{\abs{C'}}{n\cdot 2^{H(\delta) n}},\]
and thus has rate at least
\[R(C)\geq 1-H(\delta) -\frac{\e}{2}-o(1) \geq R.\]
The code $C$ is an $[n,\delta]$ binary non-linear cyclic code.
Namely, by construction, for each $x \in C$ and $i \in [n]$ it holds
that $E^i(x) \in C$.  Moreover, by the iterative procedure, any
distinct $x, x' \in C$ satisfy $d(x,x') \geq \delta$.
\end{IEEEproof}

\section{Conclusion}
\label{sec:conc}

In this work we proved the existence of a good family of binary
non-linear cyclic codes of normalized distance $\delta$ and asymptotic
rate $1-H(\delta)$ (i.e., codes that meet the GV bound). The codes we
construct are non-linear. Specifically, for $\delta<\frac{1}{2}$, the code 
$C'$ obtained in the first step of our construction is the collection of all codewords $x$ with auto-cyclic distance greater or equal to $\delta$. This code is not linear. To see this, consider any $x$ for which for all $i \in [n] \setminus \{0\}$ it holds that $d(x,E^i(x)) \geq \delta + \frac{2}{n}$. The proof of Lemma~\ref{lemma:auto} shows the existence of several such $x$. Consider now the codeword $y=x + 0^{n-1}1$. The codeword $y$ satisfies the slightly weaker condition that for all $i \in [n] \setminus \{0\}:\ d(y,E^i(y)) \geq \delta$. So both $x$ and $y$ have auto-cyclic distance at least $\delta$ however $x+y=0^{n-1}1$ has auto-cyclic distance of $\frac{2}{n}$.
The second step of our construction, which removes elements from $C'$
to obtain the cyclic code $C$ of distance $\delta$ and rate
arbitrarily close to $1-H(\delta)$, is greedy and does not necessarily
yield linear codes. Whether the first step of our construction can be
refined using algebraic techniques to yield a linear auto-cyclic code, or
whether the second step of our construction (assuming a linear $C'$)
can yield a linear code $C$, are intriguing problems left open in this
work. Problems that, if solved, may shed light on the existence of
good binary linear cyclic codes.

\section{Acknowledgment}
The authors would like to thank Alexander Barg for his valuable
comments on an earlier draft of this work.

\bibliographystyle{IEEEtranS}
\bibliography{Cyclic}

\end{document}